\documentstyle[12pt,epsfig]{article}
\begin{document}

\begin{center}
{\large{\bf Meson Spectrum and the Glueball}}

\vspace*{5mm}

\underline{G.~Ganbold$^{a,b,}$}\footnote{E--mail: {\tt ganbold@thsun1.jinr.ru}}

\vspace*{3mm}

{\it a) Bogoliubov Laboratory of Theoretical Physics, JINR, 141980, Dubna, Russia} \\
{\it b) Institute of Physics and Technology, 210651, Ulaanbaatar, Mongolia} \\
\end{center}

\vspace*{5mm}

{\small{
\centerline{\bf Abstract}
The formation and spectrum of two-particle bound states are investigated
within a simple relativistic quantum field model with the Yukawa-type
interaction. Within this approach, relatively weakly interacting quarks
and gluons form stable bound states under the analytic confinement. By
introducing a minimal set of free parameters (the quark masses, the
coupling constant and the confinement 'radius') and by solving the
Bethe-Salpeter equation with one-gluon exchange, we satisfactorily
explain the experimental data for most known mesons and the glueball.
}}
\vspace*{3mm}

\section{Introduction}

The conventional theoretical description of colorless hadrons within the
QCD implies that they are bound states of quarks and gluons considered
under the {\sl color} confinement. Another realisation of the confinement
is developed in \cite{leut81} by using an assumption that QCD vacuum is
realized by the self-dual homogeneous vacuum gluon field that can lead
to the quark and gluon confinement. Hereby, propagators of quarks and
gluons are described by entire analytic functions in the $p^2$-complex
plane \cite{efned95}, i.e. the {\sl analytic} confinement takes place,
although there exists a prejudice to the analytic confinement conception
(e.g., \cite{ahli01}).

In our earlier paper \cite{efim02} we considered simple scalar-field models
to clarify the pure role of the AC in formation of the hadron bound states.
In doing so, we have demonstrated just a mathematical sketch of calculations
of the mass spectrum for the "mesons" consisting of two "scalar quarks" by
omitting important quantum degrees of freedom such as the spin, color and
flavor. Nevertheless, these models gave a quite reasonable sight to the
underlying physical principles of the hadron formation mechanism and have
resulted in qualitative descriptions of the spectra  of "mesons", their
Regge trajectories and "glueballs" \cite{ganb04}.

It seems reasonable to consider below a more realistic extention of our
approach by taking into account the color, flavor and spin degrees
of freedom for constituent quarks and gluons. In doing so, we estimate the
spectra of pseudoscalar and vector mesons as well as the glueball lowest
state. \\

\section{Model}

Conventionally, considering this problem within QCD one deals with
complicated and elaborate calculations, because the confinement is
achieved as a result of strong interaction -- by involving high-order
corrections \cite{pros03}. Then, there arises a problem of
correct and effective summation of these contributions.

On the other hand, the use of a QFT method is effective when the
coupling is not large. Then, lower orders of a perturbative technique
can result in satisfactory accurate estimates (e.g., in QCD) of observables.
Particularly, one can effectively use the one-gluon exchange approximation.

We consider a relativistic physics and for the hadronization processes
use the Bethe-Salpeter equation, because the binding energy is not
neglible and the relativistic corrections are considerable. In
doing so, we use a minimal set of free parameters.

The Lagrangian of the Yukawa-type interaction reads:
\begin{eqnarray}
{\cal L}&=&
\left( {\bar\Psi_\alpha^i}[S^{-1}]_{\alpha\beta}^{ij}
{\Psi_\beta^j}\right) + {1\over 2}\left( \phi^a_\mu~[D^{-1}]^{ab}_{\mu\nu}
\phi^b_\nu \right)+g\left({\bar\Psi_\alpha^i}
~(i\gamma_\mu)^{\alpha\beta} t^a_{ij}~{\Psi_\beta^j}\phi^a \right) \nonumber\\
&& ~+~g\Lambda\left(\phi^a_\mu \phi^b_\nu
\left[ \partial_\mu \phi^c_\nu-\partial_\nu \phi^c_\mu \right] \right) f^{abc}\,,
\end{eqnarray}
where $\Psi_{\alpha f}^a(x)=\Psi_{{\rm spin}~{\rm flavour}}^{{\rm color}}(x)$
and $\phi_{\mu}^a(x)=\phi_{{\rm vector}}^{{\rm color}}(x)$ are the quark and
gluon fields.

We guess that the matrix elements of hadron processes at large distance
(in the confinement region) are integrated characteristics of the quark
and gluon propagators and the interaction vertices. Therefore, taking
into account the correct symmetry features is more important than the
tiny details of these functions.

Our aim is to suggest the most plain forms of these propagators
which keep the essential properties and result in a qualitative and
semi-quantative description of the hadron spectra. We consider the
following propagators for quarks and gluons:
\begin{eqnarray}
&&
\tilde{S}_{\alpha\beta}^{ij}(\hat{p})=
~{\delta^{ij}\over m^2} \left\{ i\hat{p}+m~[1+\gamma_5\omega(m)]\right\}_{\alpha\beta}
\cdot e^{-{p^2+m^2\over 2\Lambda^2}} \,, \nonumber\\
&&
D_{ab}^{\mu\nu}(x)=\delta_{ab}~\delta^{\mu\nu}~{\Lambda^2\over
(4\pi)^2}~e^{-{x^2\Lambda^2/4}} \,,
\end{eqnarray}
where $\Lambda$ - the confinement energy scale, $m$ - the quark mass for
given flavour and  $0<\omega(m)\le 1$  is a smoothly decreasing function.
These are entire analytic functions and could serve reasonable approximations
to the explicit propagators in the hadronization region.

Consider the partition function and take an explicit integration
over $\phi$-variable:
\begin{eqnarray}
Z &=& \int\!\!\!\int\delta\bar\Psi\delta\Psi
~e^{-{\cal L}_F[\bar\Psi,\Psi]} \int \delta\phi
~e^{-{\cal L}_B[\phi]-{\cal L}_{int}[\bar\Psi,\Psi,\phi]} \nonumber\\
&=& \int\!\!\!\int\delta\bar\Psi\delta\Psi
~e^{-{\cal L}_F [\bar\Psi,\Psi]-{\cal L}_2 [\bar\Psi,\Psi]
-{\cal L}_3 [\bar\Psi,\Psi] -{\cal L}_G +...} \,,
\end{eqnarray}
where the two-quark and two-gluon bound states in the one-gluon exchange
mode are generated by terms
\begin{eqnarray}
{\cal L}_2 ={g^2\over 2} \int\delta\phi ~e^{-{\cal L}_B[\phi]}
\left(\bar\Psi\Gamma\Psi\right) D \left(\bar\Psi\Gamma\Psi\right) \,,
\qquad
{\cal L}_G ={27g^2\over 2} \int\delta\phi ~e^{-{\cal L}_B[\phi]}
\left( \phi \phi D \phi \phi \right)\,.
\end{eqnarray}

\section{Light and Heavy Mesons}

Below we shortly describe important steps of our approach for two-quark
bound states. First, we allocate the one-gluon exchange between quark
currents and go to the relative co-ordinates in the centre-of-masses
system. Then, make a Fierz transformation to get the colorless bilocal
quark currents and introduce an orthonormal basis $\{U_Q(x)\}$ indiced
by the quantum numbers $Q=\{n\kappa lm\}$, where $l$ stands for the orbital
quantum number. We diagonalize ${\cal L}_2$ on colorless quark currents and
use a Gaussian representation by introducing auxialiary meson fields
$B_{\cal N}$. Then, we apply the Hadronization Ansatz to identify
$B_{\cal N}(x)$ as meson fields with quantum numbers ${\cal N}=\{JQff\}$.
The partition function for mesons reads:
\begin{eqnarray}
Z_{\cal N}=\int DB_{\cal N}~e^{-{1\over2}(B_{{\cal N}}
[1+g^2{\bf{\bf Tr}}V_{{\cal N}}SV_{{\cal N}}S]B_{{\cal N}})
+W_I[B_{{\cal N}}]}\,,
\end{eqnarray}
where $S$ - the quark propagator and
$ V_{{\cal N}}=\Gamma_J\int dy~U_Q(y)\sqrt{D(y)} \exp\{{{y\over2}\stackrel
{\leftrightarrow}{\partial}}\}$ is a vertice function.
The interaction between mesons is described by $W_I[B_{\cal N}]=O[B_{\cal N}^3]$.

The diagonalization of the quadratic form is equivalent to solution of
the Bethe-Salpeter equation (in the ladder approximation) on the orthonormal
system $U_{{\cal N}}$:
\begin{eqnarray}
g^2{\bf Tr}(V_{{\cal N}}SV_{{\cal N}'}S)
= \int\!\!\!\!\int\! dx dy~U_{JQ}(x)~g^2\Pi_p(x,y)
~U_{J'Q'}(y)=-~\delta^{QQ'}~\delta^{JJ'}~\lambda_{JQ}(-p^2) \,.
\end{eqnarray}

Then, the meson mass is defined by equation:
\begin{eqnarray}
1-\lambda_{{\cal N}}(M_{{\cal N}}^2)=0\,.
\end{eqnarray}

The polarization kernel $\Pi_p(x,y)$ is real and symmetric that allows
one to use a variational technique. For simplicity we consider orbital
excitations: $n=0$, $l\ge 0$. By solving the variational equation for
the meson mass
\begin{eqnarray*}
\label{vary}
1 \!\!\! &=& \!\!\! {\alpha_s \Lambda^4\over 3\pi m_1^2~m_2^2}
~\exp\left\{ {M_l^2(\mu_1^2+\mu_2^2)
-(m_1^2+m_2^2)\over 2\Lambda^4} \right\}~{2^l\over (l+1)!} \\
&& \!\!\!\!\! \cdot \max\limits_{0<b<1} \left\{
[b(1-b/2)]^{l+2}~(-1)^l~{d^l\over dA^l}{1\over A^2}~
\exp\left(-{M_l^2(\mu_1-\mu_2)^2\over 4A\Lambda^2}\right)
\cdot \left[ {2\rho_J\over A} \right.\right. \\
&&  \!\!\!\!\!  \left. \left. +{M_l^2(\mu_1-\mu_2)^2\over 4\Lambda^2}
\left( {2\over A} - {1\over A^2} \right)+{M_l^2\mu_1\mu_2+m_1 m_2
(1+\chi_J\cdot\omega(m_1/\Lambda)\cdot\omega(m_1/\Lambda)~)\over \Lambda^2}\right]\right\} \,, \\
&& \!\!\!\!\!
A=1+2b\,,~~~\mu_i={m_i\over m_1+m_2}\,,~~~ \rho_J=\left\{1,{1\over 2}\right\}\,,~~~
\chi_J=\left\{+1,-1\right\} ~~~ \mbox{\rm for}~~~J=\{P,V\}
\end{eqnarray*}
we fit the known experimental data and the optimal set of parameters is obtained:
\begin{eqnarray}
&& \alpha_s \equiv {g^2\over 4\pi}=0.324 \,,\qquad  \Lambda=668\mbox{\rm MeV} \,,
\qquad m_u=m_d=179\mbox{\rm MeV} \,,                            \nonumber\\
&& m_s=202\mbox{\rm MeV}  \,,\qquad m_c=781\mbox{\rm MeV}  \,,
\qquad m_b=4324\mbox{\rm MeV} \,,
\end{eqnarray}

\subsection{Meson Ground States}

With these parameters first we estimate the pseudoscalar and vector meson masses
for the ground state $l=0$ in the wide range from light $\pi(138)$ to heavy
$\Upsilon(9460)$ represented in Table 1. Note, our parameter $\alpha_s$
is close to a particular value $\alpha_s=0.34$ known from the $\tau$-lepton decay.
In the present paper we consider fixed coupling constant in the range from hundred
$\mbox{\rm MeV}$ to $\simeq \mbox{\rm 10GeV}$, although there exists another point
of view to introduce a running $\alpha_s$ dependent on the energy scale
\cite{solo02,nest03}.

\begin{table}[h]
\begin{tabular}{||c|c||c|c||c|c||c|c||}
\hline
                   &         &                       &        &                   &         &                       &        \\
  $J^{PC}=0^{-+}$  & $M_P$   &    $J^{PC}=0^{-+}$    & $M_P$  &  $J^{PC}=1^{--}$  & $M_V$   &    $J^{PC}=1^{--}$    & $M_V$  \\
                   &         &                       &        &                   &         &                       &        \\
\hline
   $\pi(140)$      &  140    &     $D(1870)$         &  1929  &   $\rho(770)$     &  770    &    $K^*(892)$    &  891   \\
   $\eta(547)$     &  547    &     $D_s(1970)$       &  2012  &   $\omega(782)$   &  782    &    $D^*(2010)$   &  2001  \\
   $\eta_c(2979)$  &  3053   &     $B(5279)$         &  5356  &   $\Phi(1019)$    &  1004   &    $D^*_s(2112)$ &  2081  \\
   $\eta_b(9300)$  &  9454   &     $B_s(5370)$       &  5398  &   $J/\Psi(3097)$  &  3097   &    $B^*(5325)$   &  5356  \\
   $K(495)$        &  495    &     $B_c(6400\pm 400)$&  6154  &  $\Upsilon(9460)$ &  9460   &                  &        \\
\hline
\end{tabular}
\caption{Estimated masses (in $\mbox{\rm MeV}$) for the pseudoscalar and vector mesons.}
\end{table}

\subsection{Meson Excites States}

Compared to the ground states, the orbital excitations $l>0$ take place
in larger distances and should be less sensitive to the short-range
details of the quark and gluon propagators. A correct description of the
mesonic Regge trajectories can serve an additional testing ground for
our model, for the approximated propagators. Our estimates (solid lines)
for the $\rho$-meson and $K$-meson families excitations plotted versus
$J=l+S$ is given in Figure 1 in comparison with the experimental data (dots).

As is expected, our model describes well the Regge trajectories of mesons.
The slope of the Regge trajectories is very sensitive on the confinement
scale parameter $\Lambda$ and its optimal value is fixed near $\Lambda=668$ MeV.

\begin{figure}[h]
 \centerline{
 \includegraphics[width=70mm,height=40mm]{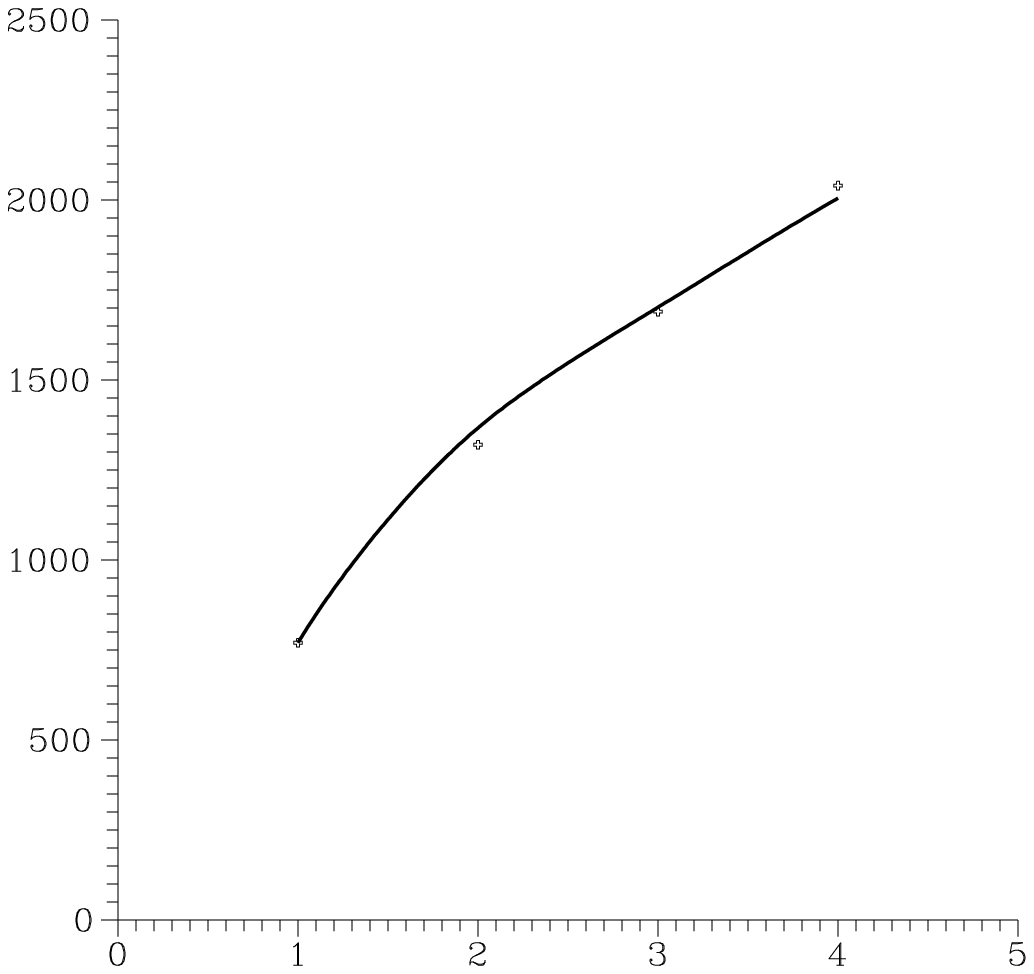}
 \includegraphics[width=70mm,height=40mm]{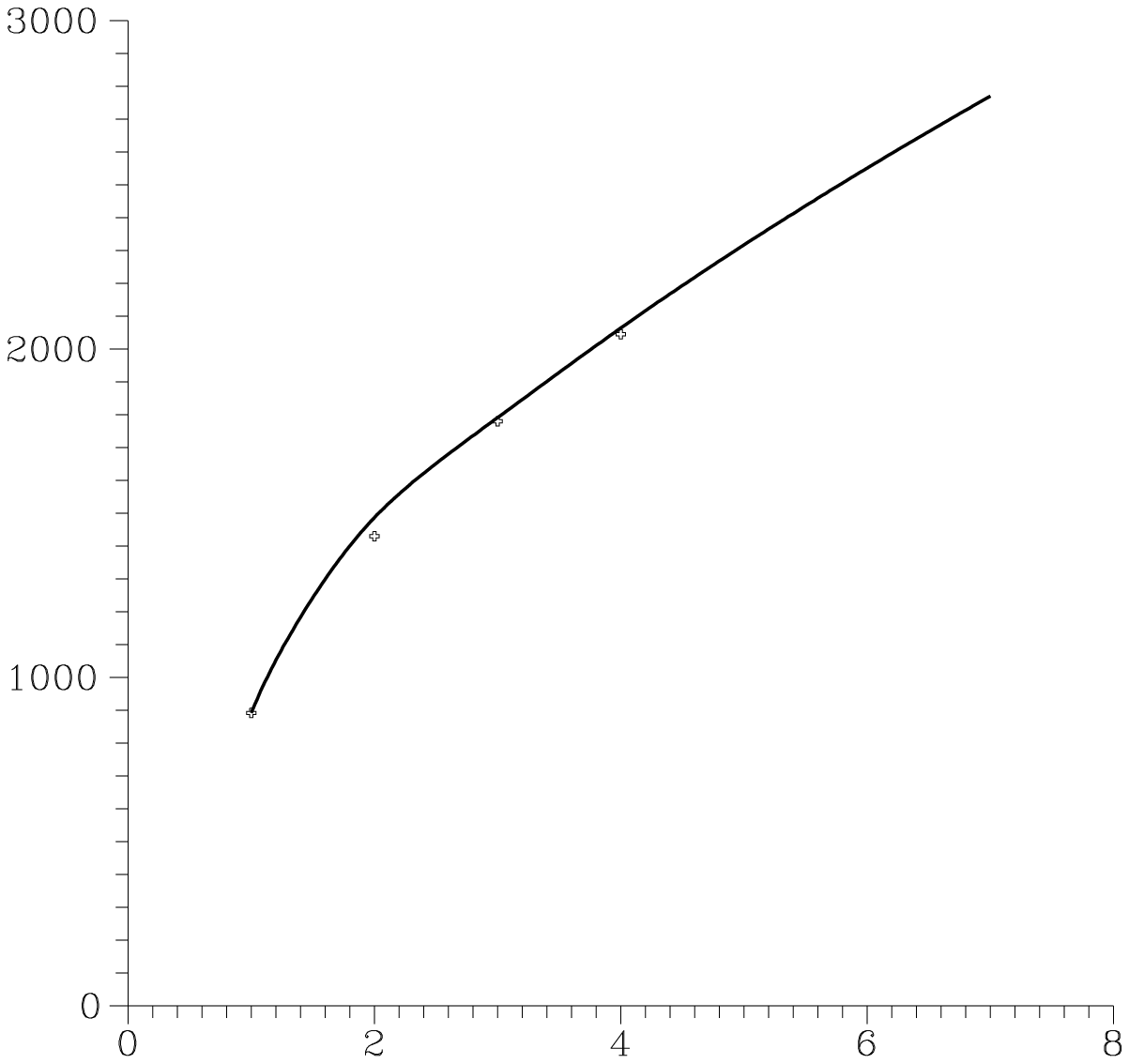} }
 \caption{
 $M_J^2$ plotted versus the quantum number $J=l+S$ (the Regge trajectories)
 for the $\rho$- and $K^*$-meson families. Solid lines correspond to our
 estimates and dots depict the experimental data from {\sl Particle Data Group 2004}.
 }
\end{figure}

\vskip 3mm

\section{Glueball}

There may exist the glueballs, bound states of gluons without any constituent
quarks. The experimental status of glueballs is not clear \cite{clem04},
although it is evident that the structure of QCD vacuum plays the main role in
their formation. We suppose that the lowest-mass glueball may be approximated by
the pure "gluon-gluon" bound state with scalar quantum numbers $J^{PC}=0^{++}$.

Omitting details, the equation for the lowest-state glueball mass reads
\begin{eqnarray}
1-\tilde{\Pi}(-M_G^2) = 0 \qquad \Rightarrow \qquad
M_G = \Lambda\sqrt{ 2\ln \left( {2\pi (2+\sqrt{3})^2 \over 27~\alpha_s} \right) }
\,.
\end{eqnarray}
With obtained values $\{\alpha_s=0.324,~\Lambda=668~\mbox{\rm MeV}\}$ we estimate
\begin{eqnarray}
M_G = 1433~\mbox{\rm MeV} \,.
\end{eqnarray}
This result may be compared with other predictions of the glueball mass:
\begin{eqnarray*}
&& M_{G_S} =1580\pm 200 MeV ~~~~~~[Y.A.Simonov, 2000]\,, \\
&& M_{G_F} =1530\pm 200 MeV ~~~~~~[H.Forkel, 2001]\,.
\end{eqnarray*}

\vskip 5mm

Concluding we remark that our simple relativistic field model with an analytic
confinement of quarks and gluons gives a quite reasonable sight to the underlying
physical principles of the hadron formation mechanism. It allows one to describe
qualitatively the meson spectra in the wide range from light $\pi(138)$ to heavy
$\Upsilon(9460)$. By using only a minimal set of fundamental parameters
(the constituent quark masses $m_f$, the coupling constant $\alpha_s$ and the
confinement scale $\Lambda$) we are able to describe the meson ground states,
the Regge trajectories of orbital excitations and the low-mass glueball state
with sufficient accuracy. Hereby, the coupling constant is relatively small
that makes the use of one-gluon exchange modes reliable for the consideration.

\vskip 5mm

The author thanks Ts.Baatar, G.V.Efimov, S.B.Gerasimov, E.Klempt, S.N.Nedelko
and W.Oelert for useful discussions and comments.



\begin{thebibliography}{99}
\bibitem{leut81} H.~Leutwyler, Phys. Lett., {\bf 96B} (1980) 154;
                 Nucl. Phys., {\bf B179} (1981) 129.
\bibitem{efned95} G.V.~Efimov and S.N.~Nedelko, Phys. Rev., {\bf D51} (1995) 174;\\
G.V.~Efimov, A.C.~Kallonaitis and S.N.~Nedelko, Phys. Rev., {\bf D59} (1999) 014026.
\bibitem{ahli01} A.~Ahlig et al., Phys. Rev., {\bf D64} (2001) 014004.
\bibitem{efim02} G.V.~Efimov and G.~Ganbold, Phys. Rev., {\bf D65} (2002) 054012.
\bibitem{ganb04} G.~Ganbold, AIP Conf. Proc., {\bf 717} (2004) 285;  {\bf 786} (2005) 127.
\bibitem{pros03} M.~Baldicchi and G.M.~Prospri, hep-ph/0310213 (2003).
\bibitem{efned96} Ja.V.~Burdanov et al., Phys. Rev., {\bf D54} (1996) 4483.
\bibitem{solo02} D.V.~Shirkov, Theor. Math. Phys., {\bf 132} (2002) 484.
\bibitem{nest03} A.V.~Nesterenko, Int. J. M. Phys., {\bf A18} (2003) 5475.
\bibitem{clem04} E.~Klempt, hep-ph/0404270 (2004).

\end{thebibliography}
\end{document}